\documentstyle[prl,aps,epsfig,floats]{revtex}

\begin{document}
\draft

\title{{\bf Sharpening the predictions of big-bang nucleosynthesis}}

\author{Scott Burles,$^1$ Kenneth M. Nollett,$^2$ James N. Truran,$^1$ and
Michael S. Turner$^{1,2,3}$}

\address{$^1$Department of Astronomy \& Astrophysics \\
Enrico Fermi Institute, The University of Chicago, Chicago, IL~~60637-1433}
\address{$^2$Department of Physics, The University of Chicago,
Chicago, IL~~60637-1433}
\address{$^3$NASA/Fermilab Astrophysics Center\\
Fermi National Accelerator Laboratory, Batavia, IL~~60510-0500}

\date{\today}

\maketitle
\begin{abstract}
Motivated by the recent measurement of the primeval abundance of deuterium,
we re-examine the nuclear inputs to big-bang nucleosynthesis (BBN).
Using Monte-Carlo realization of the nuclear cross-section data to
directly estimate the theoretical uncertainties for the yields of D,
$^3$He and $^7$Li, we show that previous estimates were a factor of
$2$ too large.  We sharpen the BBN determination of the baryon density
based upon deuterium, $\rho_B = (3.6\pm 0.4) \times 10^{-31}\,{\rm g\,
cm^{-3}}$ ($\Omega_Bh^2 = 0.019\pm 0.0024$), which leads to a
predicted $^4$He abundance, $Y_P = 0.246 \pm 0.0014$ and a
stringent limit to the equivalent number of light neutrino species:
$N_\nu < 3.20$ (all at 95\% cl).  The predicted $^7$Li abundance,
($^7$Li/H)$_P = 3.5^{+1.1}_{-0.9}\times 10^{-10}$, is higher than that
observed in pop II stars, $1.7\pm 0.3 \times 10^{-10}$ (both, 95\% cl).  We
identify key reactions and the energies where further work is needed.
\end{abstract}

\twocolumn

\section{Motivation}

Big-bang nucleosynthesis is an observational cornerstone of
the hot big-bang cosmology.  For more than two decades the predicted
abundances of the light elements D, $^3$He, $^4$He and $^7$Li have
been used to test the consistency of the hot big-bang model at very
early times ($t\sim 0.01 - 200\sec$) \cite{cst95,st98}.  The state of
affairs in 1995 was summarized by a concordance interval for the
baryon density, $\Omega_Bh^2 =$ 0.007 --  0.024, for which the
predicted abundances for all four light elements were consistent with
the observational data \cite{cst95}.  In addition to testing
the standard cosmology, BBN also gave the best determination
of the baryon density and was the linchpin in the case for
nonbaryonic dark matter.

The big-bang abundance of deuterium is
most sensitive to the baryon density \cite{reevesetal},
making it the ``baryometer.''  However, deuterium is
fragile and is destroyed by stars even before they reach the main
sequence.  Thus, local measurements of its abundance, where about 50\%
of the material has been through stars, do not directly reflect its
primeval abundance.  Recently, the situation has changed
dramatically, ushering in a new, precision era for BBN \cite{st98}.
Burles and Tytler measured the deuterium abundance in
high-redshift ($z> 3$) hydrogen clouds where almost none
of the material has been processed through stars, and they
have made a strong case for a primeval deuterium abundance,
(D/H)$_P=(3.4\pm 0.25)\times 10^{-5}$ \cite{burlestytler98a,burlestytler98b}.

From this 10\% measurement of (D/H$)_P$, the baryon density can be
inferred to around 10\%, $\Omega_Bh^2 = 0.019 \pm 0.002$, or in terms
of baryon-to-photon ratio, $\eta = (5.1\pm 0.5)\times 10^{-10}$.  With
the baryon density in hand, one can predict the abundances of the
other three light elements.  Then, $^4$He and $^7$Li can test the
consistency of BBN; D and $^3$He can probe stellar processing since
BBN; and $^7$Li can test stellar models.  Furthermore, a precise
determination of the baryon density can make BBN an even sharper probe
of particle physics (e.g., the limit to the number of light particle
species).

To take full advantage of BBN in the precision era requires accurate
predictions.  The uncertainty in the deuterium-inferred baryon density
comes in almost equal parts from the (D/H) measurement and theoretical
error in predicting the deuterium abundance.  The
BBN yields depend upon the neutron lifetime and eleven nuclear
cross sections (see Table \ref{table1}).  In 1993, Smith, Kawano and
Malaney (SKM) estimated the theoretical uncertainties by using a
conservative approach \cite{skm}.  While their
work has set the standard for the past five years, it was not without
its shortcomings: experiments were not weighted strictly by their
precision; treatment of systematic effects was neither uniform nor
explicit; fits used theoretical rules of thumb.  In addition, there
have been new measurements\cite{brune,schmid,ma}.

After a careful analysis and updating of the microphysics for small
but important effects, the theoretical uncertainty in the predicted
$^4$He abundance has been reduced to that in the neutron lifetime,
$\Delta Y_P = \pm 0.001$ (95\% cl) \cite{lopezturner98}.  Motivated by
the primeval deuterium measurement, we decided to refine the error
estimates for the other light elements, using the nuclear data
themselves and Monte-Carlo realization to make our error estimates.
This method also allowed us to identify where improvements in the
nuclear data would be most useful.

\begin{table}
\caption{For each reaction and nuclide, the energies (in keV,
center-of-mass) at which the sensitivity functions for D and $^7$Li
attain half their maximum value; these intervals indicate the energies
relevant for BBN ($\Omega_Bh^2=0.019$).}
\label{table1}
\begin{tabular}{ccc}
Reaction & {\hfil D \hfil} & 
{\hfil $^7$Li \hfil } \\
\tableline
p(n,$\gamma$)d & 25--200 & 17--153 \\
d(p,$\gamma$)$^3$He & 53--252 & 65--270 \\ 
d(d,p)$^3$H & 55--242 & 134--348 \\ 
d(d,n)$^3$He & 62--258 & 79--282 \\ 
$^3$He($\alpha$,$\gamma$)$^3$Be & no effect & 157--376 \\ 
$^3$He(d,p)$^4$He & 187--325 & 107--283  \\ 
$^3$He(n,p)$^3$H & 52--228 & 24--188 \\ 
$^7$Li(p,$\alpha$)$^4$He & no effect & 57--208 \\
$^7$Li(p,n)$^7$Be & no effect & 1649--1690 \\ 
$^3$H($\alpha$,$\gamma$)$^7$Li & no effect & 62--162 \\ 
$^3$H(d,n)$^4$He & 176--338 & 167--285 \\ 
\end{tabular}
\end{table}

\section{Method and Results}

The details of our method are described in a longer paper
\cite{burlesnollett}; here we outline the salient points.  The nuclear
inputs come in the form of measurements of cross
sections, $\sigma (E)$, or equivalently, the astrophysical $S$-factor,
$S(E)=E\sigma(E) e^{2\pi\zeta}$, where $e^{-2\pi\zeta}$ is the
Coulomb-barrier tunneling probability.  From these, the needed
thermally-averaged reaction rates per particle follow
\begin{equation}
\langle\sigma v\rangle = \sqrt{8 \over {\pi \mu (kT)^3}} \int
 \sigma(E) \, E \, e^{-E/kT} dE,
\end{equation}
where $\mu$ is the reduced mass.

\begin{figure}
\centerline{\epsfxsize=10cm \epsfig{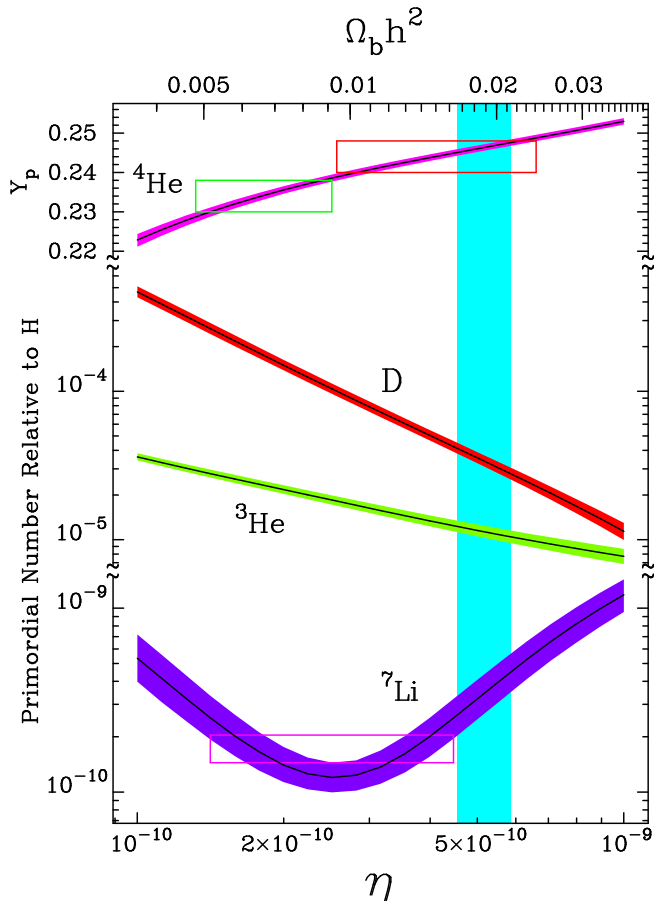}}
\caption{Summary of the 95\% confidence intervals for the BBN
predictions for D, $^3$He, $^4$He and $^7$Li.  The $^4$He
uncertainty comes from Ref.~\protect\cite{lopezturner98}.  Boxes indicate
95\% cl abundances from observation.  The vertical band indicates the
baryon density we infer from the deuterium observation.}
\label{fig:etaplot}
\end{figure}

We use Monte-Carlo realizations of all the experimental data sets to
determine thermal reaction rates and final yields.  For each
realization, we proceed as follows.  For every data point from every
data set we draw a value from a Gaussian distribution whose mean is
the central value and whose variance is the standard error reported
for that point.  We account for correlated normalization error in a
data set by similarly drawing a value for the overall normalization.
For each reaction, a smooth representation of $S(E)$ is obtained by
fitting a piecewise spline to the data, with individual points
weighted by their standard errors in the usual way.  Using
the spline fits, we evolve light-element abundances with
a standard BBN code.  From 25,000 such
realizations, we produce distributions of the light-element yields and
compute means and 95\% cl intervals.  Our results, as a function of
the baryon density, are shown in Fig.~\ref{fig:etaplot}.

Data points and uncertainties were extracted from a comprehensive
review of the experimental literature from approximately 1945 onward,
beginning with a careful reading of the original sources.  We excluded
a small number of data sets for which insufficient information for our
technique was provided.

\begin{figure}
\centerline{\epsfxsize=10cm \epsfig{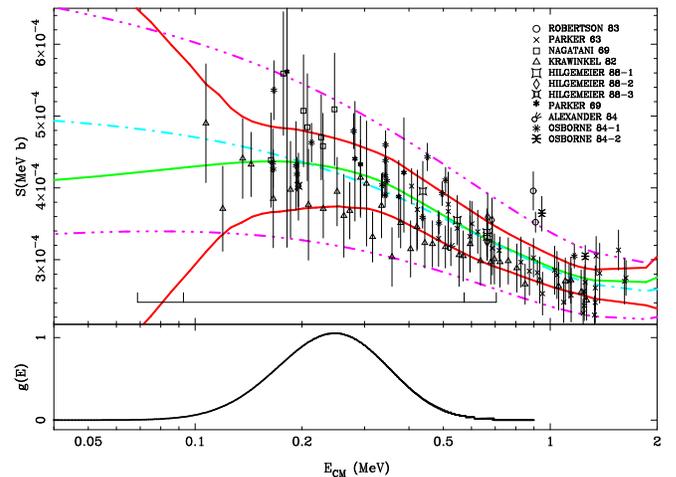}}
\caption{$S(E)$ vs. $E_{\rm CM}$ for the reaction $^3\rm{He}
(\alpha,\gamma)^7\rm{Be}$.  The data are shown with our best fit and
95\% cl interval for $S(E)$ (solid lines).  The SKM fit and 95\% cl
interval are shown as dash-dotted lines.  The integration limits on
the thermal averages needed for an accuracy of $0.1 \sigma$ (inner
tick marks) and one part in $10^5$ (outer tick marks) in the final
yields are shown.  $g(E)$ quantifies the sensitivity of the final
abundance of $^7$Li to $S(E)$; the final abundance of D is insensitive
to this reaction.}
\label{fig:he3ag}
\end{figure}

As always, there is the sticky problem of systematic error, especially
for cross sections represented by only a few measurements.  A case in
point is the reaction $^3{\rm He}(\alpha,\gamma)^7{\rm Be}$, which
produces nearly all of the $^7{\rm Li}$ for $\eta=5.1 \times
10^{-10}$.  Activation measurements \cite{robertson,osborne,volk} show
an apparent disagreement with prompt-photon measurements (see
Fig. \ref{fig:he3ag} and Ref.~\cite{adelberger}).  Because these
measurements are not in the energy range of relevance for BBN, they
have little influence on our results.  (SKM omitted activation
measurements from their analysis altogether.)  We take them into
account by performing a second Monte Carlo, where the prompt-photon
measurements are renormalized by the weighted mean (and uncertainty)
of the three activation measurements.  This shifts the $^7$Li/H 95\%
cl interval upward by 11\% (see Fig.~\ref{fig:li7}).

Our method breaks down completely for the process $p+n\rightarrow d +
\gamma$.  This is because of a near-complete lack of data at the
energies relevant for BBN.  The approach used for this reaction is a
constrained theoretical model that is normalized to high-precision
thermal neutron capture cross-section measurements.  In particular, we
use the most recent evaluation, from ENDF-B/VI \cite{png}.  This
evaluation was performed around 1970 (with a minor update in 1989),
and it fitted a capture model to data of similar vintage for the
neutron-proton system.  No documentation survives, and the uncertainty
is difficult to quantify --- especially in light of known systematic
problems with the likely input data \cite{hale}.  (Efforts are
underway to construct a new model for this reaction, based upon more
modern nuclear models and data \cite{nollettetal99}.)  For
consistency, we follow SKM and assign a 5\% $1\sigma$ uncertainty in
the overall normalization (also consistent with an estimate from the
evaluation's authors \cite{hale}), and we use this value for our
Monte-Carlo calculations.

\begin{figure}
\centerline{\epsfxsize=10cm \epsfig{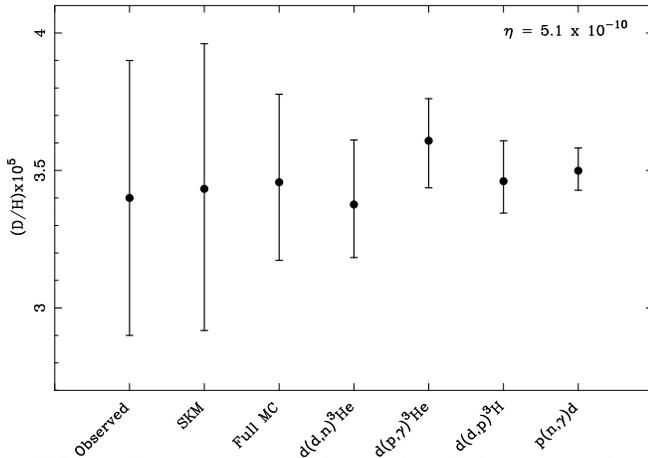}}
\caption{Uncertainties in the predicted deuterium abundance from SKM,
our full Monte Carlo, and individual reactions, compared with the
Burles \& Tytler\protect\cite{burlestytler98b} measurement.  The
uncertainties due to reactions not shown are not important.}
\label{fig:deuterium}
\end{figure}

\begin{figure}[t]
\centerline{\epsfxsize=10cm \epsfig{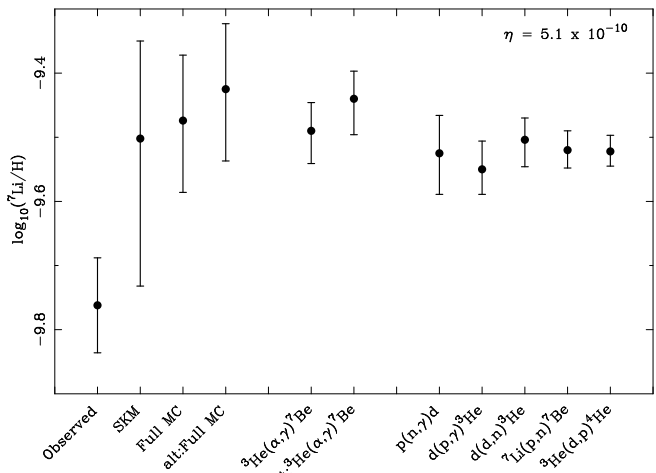}}
\caption{Same as Fig.~\protect\ref{fig:deuterium}, but for $^7$Li.
The results for the alternative normalization of $^3$He$(\alpha
,\gamma )^7{\rm Be}$ are also shown.  }
\label{fig:li7}
\end{figure}

To investigate the role of each reaction independently, we ran the BBN
code using the SKM rates for all but one reaction, studying that
reaction alone with our Monte-Carlo method.  This produced, for each
of the eleven key reactions, a best fit to the cross-section data,
95\% cl intervals for the cross sections (Fig.~\ref{fig:he3ag}),
and 95\% cl uncertainties for D and $^7$Li yields for each reaction
(see Figs.~\ref{fig:deuterium} and \ref{fig:li7}).

Our most important result is apparent: the uncertainty estimate from
our method is a factor of
two smaller than the SKM estimate.  Not only have we reduced the
theoretical error estimate, but we have also put it on a firmer
footing.  Our ``most probable'' yields also differ slightly (less than
$1\sigma$) from the corresponding results of SKM.  This reflects both
differences in weighting the nuclear data, and the inclusion of new
data.

We computed ``sensitivity functions'' for the yields of D and $^7$Li
for each reaction.  These functions measure the fractional changes in
yield caused by a delta-function change in cross section at a given
reaction energy (see Fig.~\ref{fig:he3ag} and Table \ref{table1}).
The sensitivity functions quantify where precise cross section
measurements are required.

\section{discussion and conclusions}

We have reduced the theoretical error estimate for BBN deuterium
production by a factor of two, so that the deuterium abundance itself
dominates the uncertainty in baryon density.  The deuterium
determination of the baryon density is thus sharpened, from 8\% to
6\% (at $1\sigma$), or $\Omega_Bh^2 = 0.019\pm 0.0024$ (95\% cl).
In the next five years, the
precision of the primeval deuterium measurement should improve
significantly, because the Sloan Digital Sky Survey will increase the
number of QSOs with measured redshifts by a factor of almost 100, with
a similar increase in the number of deuterium systems expected.
Further improvement in the theoretical prediction is possible; the key
reactions in this regard are: d(p,$\gamma$)$^3$He; d(d,p)$^3$H above
100\,keV; d(d,n)$^3$He above 100\,keV; and p(n,$\gamma$)d at
30--130\,keV (see Fig.~\ref{fig:deuterium}).  Turning the deuterium
determination of the baryon density into a few percent measurement
will make possible a beautiful consistency test \cite{st98}:
comparison with a similarly accurate measurement of the baryon density
from CBR anisotropy.

The deuterium-inferred baryon density leads to a prediction for the
big-bang $^4$He abundance: $Y_P = 0.246 \pm 0.001 \,({\rm D/H}) \pm
0.001\, (\tau_n )= 0.246\pm 0.0014$ (all 95\% cl).  When the primeval
$^4$He abundance is determined to three significant figures, this will
be a powerful consistency test.  At the moment, systematic effects
dominate the error budget; in particular, underlying stellar
absorption in the most metal-poor H{\sc ii} regions.  Izotov and
Thuan's sample \cite{it98,it97} excludes the tainted or
suspected-to-be tainted systems, and they find $Y_P=0.244\pm 0.002$.
This is consistent with the deuterium prediction.  A less homogeneous
sample \cite{oss97}, which includes some of the tainted systems,
indicates a lower value, $Y_P =0.234\pm 0.002$, which is not
consistent with the deuterium prediction.

Additional light particle species present around the time of BBN lead
to increased $^4$He production, and an upper limit to the primeval
$^4$He abundance can be used to constrain their existence \cite{ssg}.
Using $Y_P=0.244\pm 0.002$, the deuterium-determined baryon density,
and the prior $N_\nu \ge 3.0$, we derive the 95\% cl limit, $N_\nu
< 3.20$.  One should be mindful that systematic error in $Y_P$ could
change the limit, and that it will become more secure with better
$^4$He measurements.

Finally, we turn to $^7$Li, the light element for which the
uncertainty in the predicted abundance is largest.  Our analysis has
reduced the theoretical uncertainty by a factor of two, though a small
systematic uncertainty remains.  Using our full Monte Carlo with the
deuterium observations, we predict ($^7$Li/H)$_P= [3.5^{+1.1}_{-0.9} +
0.4\ ({\rm sys})] \times 10^{-10}$. The abundance derived from old, pop II
halo stars is ($^7$Li/H)$_{\rm pop\ II} = [1.73\pm 0.1\ {\rm (stat)}
\pm 0.2\ {\rm (sys)}]\times 10^{-10} = (1.73\pm 0.3)\times 10^{-10}$
\cite{bono} (all at 95\% cl). The discrepancy could represent a real
inconsistency or merely a depletion of $^7$Li by a factor of around
two in these stars (predicted by some models of stellar evolution
\cite{depleteLi7,pinso98}).  A nuclear solution for the discrepancy
is unlikely --- a $\sim 25\%$ (or $5\sigma$) change in the
$p(n,\gamma)$D rate would be required, and the unresolved systematics
of $^3{\rm He}(\alpha,\gamma)^7{\rm Be}$ can only make the problem
worse.  There is still much room to improve the BBN $^7$Li prediction;
the key reactions are: $p(n,\gamma)d$; $^3{\rm
He}(\alpha,\gamma)^7{\rm Be}$; $d(d,n)^3{\rm He}$; and
$d(p,\gamma)^3{\rm He}$.

Perhaps the most rewarding result of this work is that we have
verified what David Schramm many times proclaimed, ``the
predictions of BBN are very robust because the key cross section are
measured at the energies where they are needed.''  In particular, if
all eleven critical cross sections were set to zero outside the
intervals where they are measured, the final light-element abundances
would change by less than 10\% of their current theoretical uncertainty.

\acknowledgments This project was initiated and inspired by David
N. Schramm, and we dedicate this paper to his memory.  




\end{document}